# The Enhancement of Confocal Images of Tissues at Bulk Optical Immersion

I. V. Meglinski[1], A. N. Bashkatov[2], E. A. Genina[2], D. Y. Churmakov[1], and V. V. Tuchin[2]

[1] *School of Mechanical Engineering, Cranfield University, Cranfield, MK43 0AL, UK*
e-mail: i.meglinski@cranfield.ac.uk

[2] *Department of Physics, Saratov State University, Saratov, 410026, Russia*
e-mail: tuchin@sgu.ru@

**Abstract**—The purpose of the present work is a theoretical examination of how localized skin-tissue dehydration affects the depth of the confocal probing and what depth of effective detection can be reached with the chemical administration of skin tissues. A semi-infinite multilayer Monte Carlo model is used to estimate spatial localization of the output signal offered by a confocal probe. A solution of glycerol is taken in the capacity of innocuous osmotic agent. Diffusion of this bio-compatible chemical agent into the skin temporarily pushes water out of the tissues and results in the matching of the refractive indices of skin structural elements. This temporarily decreases scattering and increases transparency of topical skin layers, which allows for unrestricted light to permeate deeper into the skin. The results of simulation show that signal spatial localization offered by a confocal probe in the skin tissues during their clearing is usable for the monitoring of deep reticular dermis and improving the image contrast and spatial resolution. A discussion of the optical properties of skin tissues and their changes due to diffusion of glycerol into the skin is given. Optical properties of tissues and their changes due to chemical administration are estimated based on the results of experimental in vitro study with rat and human skin.

## 1. INTRODUCTION

The unique characteristics of optical/laser methods are of great interest to researchers working in various areas of biology and medicine [1, 2]. Nevertheless, the problem of implementing these techniques in clinical practice in order to solve a wide range of actual diagnostic tasks remains unresolved. The difficulties in clinical application are due to highly anisotropic scattering of the probing laser radiation in most biological tissues. Furthermore, random inhomogeneous variations of the optical properties of the topical skin layers act like a screen, which keeps stray optical radiation from penetrating deeply into the human body [3]. The mathematical expression for the propagation of optical radiation describing these conditions is complex. Practically, it is extremely difficult to distinguish regular waves corresponding to internal structure of the medium or individual characteristics of the scattering particles.

Diffusive injection of some chemical agents in biological tissues results in reduced light scattering [4–6]. The preliminary experiments on rat skin [5, 7–9], where glucose, glycerol, trazograph, polyethylen glycol, cosmetic lotions, and gels were taken in the capacity of innocuous bio-compatible chemical agents, showed that diffusion of the chemicals into skin temporary pushes water out of the upper skin tissues. This causes matching of the refractive indices of the structural elements of skin cells and temporarily increases the transparency of upper skin layers, which in turn allows the unrestricted light to permeate deeper into tissues. The augment of the transparency of skin layers typically increases for 10–40 minutes, depending on the chemical agent. Then, due to the physiological response of the organism, water content of skin tends back to its initial level, and the transparent skin returns to normal. Nevertheless, this temporal gap of scattering reduction can significantly improve the efficiency of skin image reconstruction techniques, such as confocal microscopy or optical coherence tomography (OCT), for instance, and provide information that is not available for dermatologists and other researchers at the present time. However, an assessment of the depth to which we can penetrate into the skin by optical probing during administration of chemicals to skin tissues is required.

The goal of the present work is to examine how the localized dehydration of skin tissues affects the depth of confocal probing and what effective detection depth can be reached with such chemical administration of skin tissues.

## 2. MATERIALS AND METHODS

Stochastic numerical modeling is used to study the spatial localization of the output signal of a confocal probe within human skin tissues. The basis for this technique is the Monte Carlo (MC) approach [10, 11], which excludes the energy conservation problem inherent to the previous MC algorithms [12–14]. The details of the technique are justified below. The confocal probing is simulated following the ray-tracing approach coupled with MC [15] for a multilayer medium. The layers correspond to the randomly inhomogeneous distribution of blood, chromophores, and various pigments within skin tissues [16–18] that produce spatial variation of their optical properties, i.e., scattering coefficient $\mu_s$, absorption coefficient $\mu_a$, anisotropy factor $g$, and refractive index $n$ [1]. Boundaries of the layers are simulated as wavy surfaces (Fig.1) [10, 11] with respect to the anatomical peculiarities of the structure of human skin cells, as well as the spatial distribution of chromophores, pigments, and blood vessels [16–18].

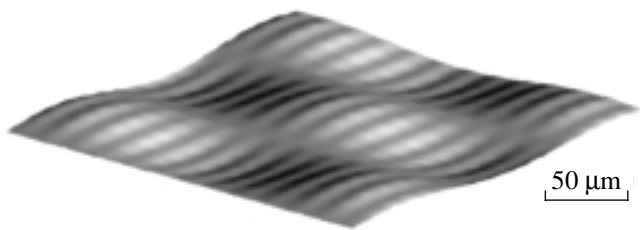

**Fig. 1.** An example of a randomly wavy periodic surface representing the interfaces between the derma layers [10, 11].

We subdivide the epidermis into two sublayers called nonliving and living epidermis. Nonliving epidermis, or stratum corneum (about 20 μm thick), consists of only dead squamous cells, which are highly keratinized with a high lipid and protein content, and has a relatively low water content [17]. This content defines the density of average absorption and scattering cross-sections for the layer. Living epidermis (~100 μm thick) contains most of the skin pigmentation, mainly melanin, which is produced in the melanocytes. Large melanin particles such as melanosomes (>300 nm in diameter) exhibit mainly forward scattering. In contrast, melanin dust, whose particles are small (<30 nm in diameter), has anisotropy in the scattering profile, and the optical properties of melanin particles (30–300 nm in diameter) can be predicted by the Mie theory. Random spatial distribution of melanin particles in living epidermis affects both the scattering and absorption properties of human skin [19].

Derma is a vascularised layer, and the main absorbers are the blood-bone pigments hemoglobin, carotene, and bilirubin. The optical properties of derma, mainly absorption, depend on the distribution of blood containing haemoglobin. Following the distribution of blood vessels in skin [16, 18] we subdivide dermis into four layers: papillary dermis (~150 μm thick), upper blood net plexus (~100 μm thick), reticular dermis (~1150 μm thick) and deep blood net plexus (~100 μm thick). Scattering properties of dermal layers are mainly defined by the fibrous structure of the tissue, where collagen fibers are packed in collagen bundles and have lamellae structure [16]. The light scatters on both single fibers and scattering centers, which are formed by the interplacement of collagen fibers.

The subcutaneous fat (~6000 μm thick) is formed by aggregation of white adipocytes fat cells (~120 μm in diameter) containing stored lipid oil in the form of single droplets of triglyceride.

Table 1 outlines the summarized optical properties of skin layers. These properties correspond to the intrinsic optical parameters of normal human skin (Caucasian type) at 633-nm wavelength. The refractive index of external medium is 1.0 (air), whereas for the deepest area it is assumed to be the same as for subcutaneous fat.

The diffusion of exogenous chromophores or biocompatible chemical agents into the skin results in the matching of refractive indices of scattering elements of skin tissues and interstitial fluid of its tissue [1]. Thereupon, the path-length in a forward direction for the incident radiation increases as the unrestriction of incident light on the refractive-index mismatching boundaries decreases. This plays a significant role for the shallow optical probes [10, 20].

It should be pointed out that direct measurements of the optical properties of the skin layers described above are impossible. We estimate the changes in optical properties of the skin layers based on their chromophores content and micro structure [16–18] typical for normal human skin.

Difference in blood and lymph-vessel distribution, cell structure, and packing of the collagen fiber bundles within the skin produces complex oscillations in the diffusion of glycerol ($n = 1.45$) and other agents within the skin tissues. We assume that the scattering properties of blood-containing layers become nonlinear due to blood flow (and also due to lymph flow), which periodically washes away a part of an osmotic agent. In the model, we also emphasize that the dermal–epidermal junction and variations of size and packing of topical skin cells and collagen fiber bundles produce spatial variations in the glycerol concentration. This is represented schematically in Fig. 2. Stratum corneum greatly clears in the first minute of the process [8, 21]. It is connected mainly with an immersion of dead cells just after administering the agent. Then, diffusion is dramatically reduced due to the barrier produced by the

Optical coefficients of the skin layers used in the simulation

| | Skin layer | $\mu_s$, mm$^{-1}$ | $\mu_a$, mm$^{-1}$ | $g$ | $n$ |
|---|---|---|---|---|---|
| 1 | Stratum corneum | 80 | 0.1 | 0.8 | 1.53 |
| 2 | Living epidermis | 35 | 0.15 | 0.8 | 1.34 |
| 3 | Papillary dermis | 30 | 0.07 | 0.9 | 1.4 |
| 4 | Upper blood net dermis | 25 | 0.01 | 0.95 | 1.39 |
| 5 | Reticular dermis | 20 | 0.07 | 0.76 | 1.4 |
| 6 | Deep blood net dermis | 30 | 0.12 | 0.95 | 1.39 |
| 7 | Subcutaneous fat | 15 | 0.07 | 0.8 | 1.44 |

Note: The refractive index of outside ambient (air) is $n_0 = 1$.

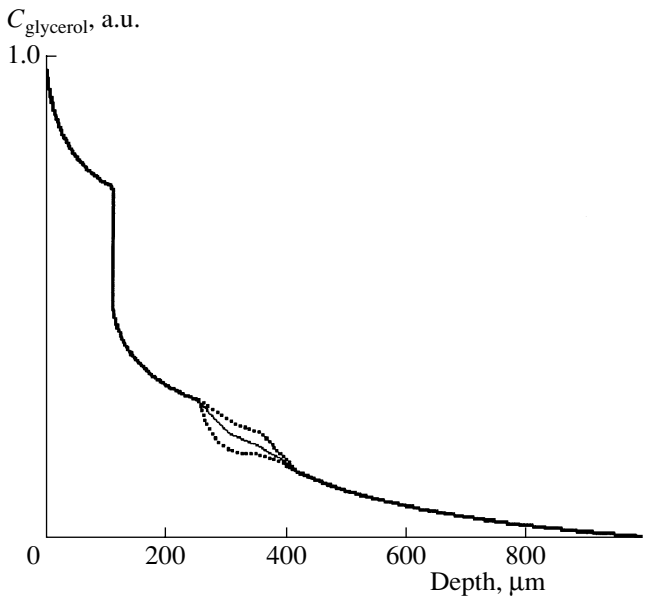

**Fig. 2.** Schematic profile of glycerol concentration changes in the outer layers of skin. Dots represent the variations of glycerol concentration in upper blood net dermis due to blood flow.

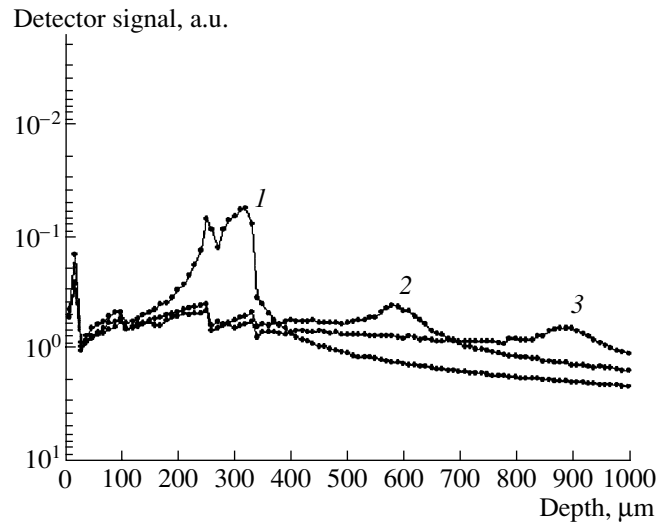

**Fig. 3.** The axial profile of detector-signal distribution predicted by the numerical MC simulation for a confocal probe focusing at (*1*) 300 μm, (*2*) 600 μm, and (*3*) 900 μm into the tissues. Confocal probe parameters are as follows: lens diameter, 5 mm; focal length, 10 mm; pinhole diameter, 10 μm; height of the lens above the surface, 9.7 mm. Optical properties of skin layers are presented in Table 1.

dermal–epidermal junction (see Fig. 2). Intra-dermal diffusion of chemical agents is characterized by retarding in the upper blood net plexus that seems to be explained by blood and lymph vascular net distribution [16, 18, 22]. In deep skin layers, the character of glycerol diffusion is more homogeneous and significantly slower. We estimate the average value of glycerol diffusion coefficient in the reticular dermis as $7.5 \times 10^{-5}$ mm$^2$/sec at 36.9°C and its variations in the layers as $0.18–0.25 \times 10^{-5}$ mm$^2$/sec. These data are evaluated based on the model of free diffusion, the Gladstone and Dale law [1], and experimental results [8, 21].

Current model was used to calculate axial focusing profiles of the detecting signal distribution [11, 15] for a few values of confocal focusing. The obtained change in the detection-signal distributions is proportionally related to the effective path-length distributions through the tissue and/or to the glycerol concentration changes. Profiles of the detection signal are simulated separately for each focusing depth and stage of glycerol diffusion, since it depends on the optical properties of tissues and probing geometry.

## 3. RESULTS

Figure 3 presents simulated profiles of the detected signal distributions for a confocal probe focusing at 300, 600, and 900 μm in a modeling medium. The parameters of the probe used in the simulation are as follows: lens diameter, 5 mm; focal length, 10 mm; pinhole diameter, 10 μm. Optical properties corresponding to the properties of skin layers are presented in Table 1. The results show (see Fig. 3) that at the normal stage of skin, the main signal is collected from its topical area (20 μm thick). This illustrates the screening effect of skin layers and how quickly the topical skin layers keep stray optical radiation from deeply penetrating into the deep tissues. Nevertheless, for shallow focusing of the confocal probe (at 300 μm), a significant part of the detecting area with a total thickness of about 150 μm is localized around the focal plane.

Optical changes of topical skin layers by the glycerol diffusion distort the screening effect. Decreasing of scattering of topical skin layers ($\mu_s^{(1)} = 20$ mm$^{-1}$, $\mu_s^{(2)} = 15$ mm$^{-1}$, $\mu_s^{(3)} = 25$ mm$^{-1}$, $n_1 = 1.45$, $n_2 = 1.4$, other optical properties are unchanged; see Table 1) significantly increases the detection signal at 600 μm focusing depth (see Fig.4). The main detected signal is collected from a middle part of dermis, but the influence of epidermis and topical layers of dermis is still significant. We ascribe these optical changes to the 10 minutes of topical-glycerol diffusion into the skin tissues. Posterior matching of refractive indices (20 minutes later) substantially increases the localization of signal at 900 μm and even deeper (see Fig. 5). These results are presented for $\mu_s^{(1)} = 15$ mm$^{-1}$, $\mu_s^{(2)} = 10$ mm$^{-1}$, $\mu_s^{(3)} = 10$ mm$^{-1}$, $\mu_s^{(4)} = 12$ mm$^{-1}$, $\mu_s^{(5)} = 15$ mm$^{-1}$, $n_{1-5} = 1.4$, and other optical properties are unchanged (see Table 1). The influence of topical layers of dermis is decreased (see Fig. 4 and Fig. 5 to compare). Spatially, the detector signal offered by a confocal probe is localized mainly around the focal plane area for all depths of focusing (see Fig. 5).

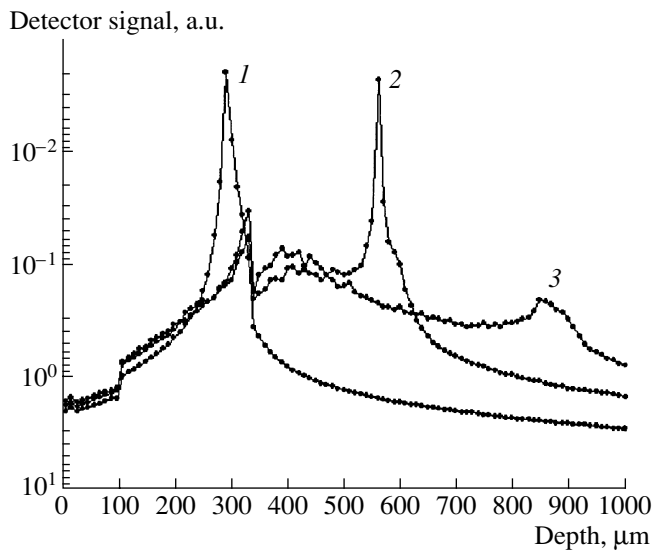

**Fig. 4.** The axial profile of a detector-signal distribution predicted by numerical MC simulation for a confocal probe focusing at (1) 300 μm, (2) 600 μm, and (3) 900 μm into the skin after 10 minutes of glycerol diffusion. Confocal probe parameters are as follows: lens diameter, 5 mm; focal length, 10 mm; pinhole diameter, 10 μm; height of the lens above the surface, 9.7 mm. Optical properties of skin layers are presented in Table 1 and discussed in the text.

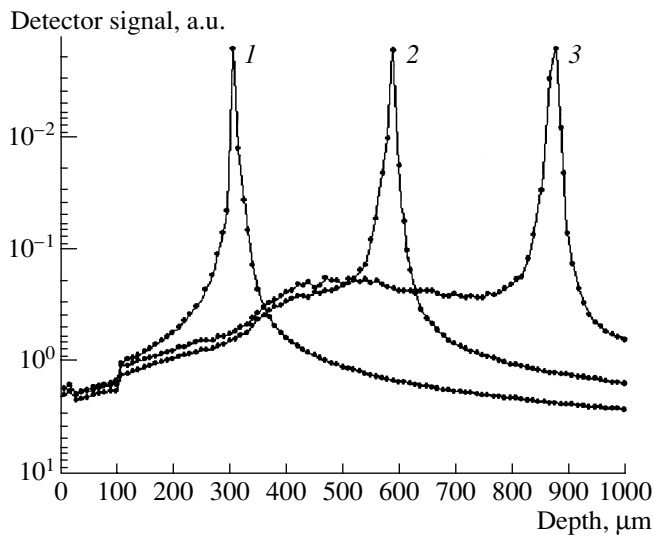

**Fig. 5.** The axial profile of a detector signal distribution predicted by numerical MC simulation for a confocal probe focusing at (*1*) 300 μm, (*2*) 600 μm, and (*3*) 900 μm into the skin at final maximal stage of skin clearing, after 20 minutes of glycerol diffusion. Confocal probe parameters are as follows: lens diameter, 5 mm; focal length, 10 mm; pinhole diameter, 10 μm; height of the lens above the surface, 9.7 mm. Optical properties of skin layers are presented in Table 1 and discussed in the text.

## 4. DISCUSSIONS AND CONCLUSIONS

Using a simple numerical method, we have estimated what depth is preferentially sensitive for a confocal probing during clearing of skin tissue by glycerol. The results of the simulation show that the spatial localization offered by a confocal probe in skin due to its clearing is potentially usable for monitoring upper dermal layers, i.e., three times deeper than original probing depth. The present computational model could not consider the dynamics of the changes in skin-tissue optical properties, which is an important factor. Nevertheless, in the framework of the simulation, we have estimated the optical properties of skin layers based on experimental measurements of the glycerol diffusion of skin tissue [1, 8, 21] and taken into account the observed structure of skin tissues. We emphasize that hypodermal injection of bio-compatible chemical agents increases the speed of the tissue clearing, since this agent diffuses straight into the dermal layers with the exception of the barriers, whereas, for topical superficial administering of skin tissues, the diffusion processes are slow, and the time of clearing is significantly (~10 times) longer. We will present these data in future reports upon final establishment of the geometry changes of skin layers.

The obtained results can be useful to evaluate the capabilities of skin-imaging systems based on confocal probing. The current approach has the potential to dramatically improve the specificity of visual diagnostic techniques such as colposcopy, fluorescence microscopy, optical coherence tomography (OCT), and other medical diagnostic techniques and could be useful in laser surgery as well.

## ACKNOWLEDGMENTS

I.V.M. and D.Y.C. acknowledge the support of the Department of Optical and Automotive Engineering, School of Engineering, Cranfield University, and A.N.B., G.E.A., and V.V.T. acknowledge the financial support of the Russian Foundation for Basic Research, program "Leading Scientific Schools," grant no. 00–15–96667, and award no. REC–006 of the U.S. Civilian Research and Development Foundation for the Independent States of the Former Soviet Union.